\documentclass[aps,showpacs,twocolumn]{revtex4}%
\usepackage{amsfonts}%
\usepackage{amsmath}%
\usepackage{amssymb}%
\usepackage{graphicx}

\begin{document}
\preprint{}
\title[The structure of the ground superfluid state...]{The structure of the ground superfluid state in a gas of Fermi atoms near the Feshbach resonance}
\author{Yu. Kagan}
\author{ L. A. Maksimov}
\affiliation{Kurchatov Institute, Moscow 123182, Russia}

\begin{abstract}
Within the framework of the variational approach the ground state
is studied in a gas of Fermi atoms near the Feshbach resonance at
negative scattering length. The structure of the originating
superfluid state is formed by two coherently bound subsystems. One
subsystem is that of quasi molecules in the closed channel and the
other is a system of pairs of atoms in the open channel. The set
of equations derived allows us to describe the properties of the
ground state at an arbitrary magnitude of the parameters. In
particular, it allows one to find a gap in the spectrum of
single-particle Fermi excitations and sound velocity
characterizing a branch of collective Bose excitations.
\end{abstract}
%\volumeyear{} \volumenumber{} \issuenumber{}
%%\eid{identifier}
%\date[Date text]{}
%\received[Received text]{}

%\revised[Revised text]{}

%\accepted[Accepted text]{}

%\published[Published text]{}

\startpage{1}
\endpage{}
\maketitle

Recently, the trend of investigations, associated with ultracold
gases of Fermi atoms and Bose molecules generated on their base,
has been established and actively developed. The key point here is
a use of the Feshbach resonance for the pair interaction of
low-energy Fermi particles. The dependence of the resonance
position on steady magnetic field $B$ has given a unique
opportunity for continuous variation of the effective
interparticle interaction and for the change its sign (sign of
scattering length $a$) with variation of magnetic field $Â$
\cite{1}.

Several stages of studies in that field  elapsed for a relatively
short time interval. At the first stage there is demonstrated  a
production of ultracold molecules $^{40}K_{2}$ \cite{2} and
$^{6}Li_{2}$ \cite{3}, \cite{4}, \cite{5} in the vicinity of the
Feshbach resonance at magnitudes $B$ corresponding to the positive
scattering length and, thus, existence of the genuine bound state
near the dissociation threshold. In both cases the resonance
occurs in the channel of $S$-scattering. For fermions polarized
over electron spin, this supposes the interaction of non-identical
atoms in different hyperfine states.

It is found that the molecules produced live relatively long in
such excited state. This supposes a possibility for the quasi
adiabatic transition with variation of field $B$ from the state
with $a<0$ in which genuine molecules do not form into the state
with $a>0$ and {\it vice versa}. The nature for suppression of
inelastic processes during collisions with the transition of
molecules to deeper levels is explained consistently in \cite{6}.

The conservation of molecules at one vibrational level facilitates
the conditions for Bose-Einstein condensation (BEC) in originating
gas of composite bosons. In the explicit form the condensation of
excited molecules is discovered in a gas of atoms $^{40}K\
$\cite{7} as well as in $^{6}Li$ \cite{8}, \cite{9}, \cite{10}.

The greatest interest in this field is connected with the study of
Fermi atoms near the Feshbach resonance at magnetic fields
corresponding to the scattering length $a<0$.

A possibility to choose the parameters, converting rarefied gas
into the system of strongly interacting particles with the
simultaneous lack of real molecules, makes an analysis of
many-particle superfluid state, which appears in these conditions,
especially interesting. For the first time, the question on the
use of the Feshbach resonance for achieving superfluidity in a
ultracold gas of Fermi atoms is discussed in \cite{11}, \cite{12}.
The problem of reconstructing the superfluid state  in the course
of the transition from the weak attraction to the strong one is
considered earlier in the known papers \cite{13}, \cite{14}. The
interaction had a single channel character and the spin
configuration remained unchanged. The specific feature of the
Feshbach resonance lies in the two-channel character of the
interaction. Particles in the continuous spectrum interact via the
formation of the intermediate quasi-molecule state in the other
spin configuration. A snapshot fixes a two-component character of
the system with pairs of particles in the quasi-molecular or
unbound states. A dynamic coherent exchange realizes between  two
subsystems, providing permanency in equilibrium at the amplitude
level of the population for the quasi-molecular state (decaying in
vacuum). The coherent superposition of the quasi-molecular states
and pairs of atoms in the continuous spectrum provides the
formation of the superfluid state with the unified condensate.

For the first time, the picture of such state is revealed
experimentally at first in gas $^{40}K$ \cite{15} and then in gas
$^{6}Li$ \cite{16}. The idea of experiment is in the adiabatic
projection of the $a<0$ state onto the genuine molecular state
with $a>0$ due to sufficiently rapid variation of field $B$,
preventing from the formation of the condensate in the course of
the transition. Special measurements have confirmed the latter
assumption, see, in particular, recent paper \cite{17}. The
experimental results in this field have stimulated the appearance
of a series of theoretical works
\cite{18},\cite{19},\cite{20},\cite{21},\cite{22} with the attempt
of the microscopic description of the strongly interacting Fermi
atoms, employing, however, one or another variants of the
perturbation theory.

The present work is devoted to finding the ground state of the
interacting Fermi atom system near the Feshbach resonance at $a<0$
within the framework of variational approach. As a result, there
appears an opportunity for unified description of  the ground
state at an arbitrary relation of parameters.

A choice of the macroscopic variational function $\Psi_{0}$
reflects the fact that the ground state of the system in the
vicinity of the Feshbach resonance is a unified state formed in
essence by two coherently bound subsystems, namely, subsystem of
Fermi atoms in the BCS-like state and subsystem of Bose
quasi-molecules in the condensate state. In addition,  $\Psi_{0}$
contains the states of quasi-molecules with nonzero momentum of
the gravity center $k\neq0\ $ and those of pairs of atoms with
nonzero total momentum. This allows from the very beginning to
involve  abovecondensate excited states (temperature $T=0$), which
play a noticeable role during the formation of the ground state,
into the variational procedure. The system of equations obtained
gives an opportunity, in particular, to find a gap in the spectrum
of single particle excitations and velocity of sound
characterizing a branch of collective Bose excitations and, thus,
to determine the critical velocity.

In the present work we give a solution of this set of equations in
explicit form for the limiting cases which allow an analytical
solution. We restrict ourselves with an assumption of the
smallness of the resonance coupling constant. The numerical
solution in the general case will be given elsewhere.

To make analysis more transparent, we consider the uniform system.
The exchange between the subsystems of Fermi atoms and
quasi-molecules with the conservation of the total number of atoms
determines the appearance of a single chemical potential $\mu$.
The generalized Hamiltonian of the system can be written as
\begin{equation}
\hat{H}^{\prime}=\hat{H}-\mu\hat{N}=\sum_{p,\sigma}\xi_{p}\hat{a}_{p\sigma
}^{+}\hat{a}_{p\sigma}+\sum_{k}\zeta_{k}\hat{b}_{k}^{+}\hat{b}_{k}+\hat
{V}+\hat{V}_{a}^{\prime}+\hat{V}_{a}^{\prime\prime}. \label{1}%
\end{equation}
Here $\hat{a}_{p\sigma}$ is the annihilation operator of Fermi
atom in the open channel, $\sigma=1,2$ identifies the hyperfine
state of an atom in the binary mixture and $\hat{b}_{k}$ is the
annihilation operator of quasi-molecules.
\[
\xi_{p}=\hbar^{2}p^{2}/2m-\mu,\ \ \zeta_{k}=\zeta_{0}+\hbar^{2}k^{2}%
/4m,\ \ \zeta_{0}=\varepsilon_{0}-2\mu,
\]
where $\varepsilon_{0}$ is a position of the resonance level in
the closed channel, $m$ is the atom mass (we neglect a difference
between hyperfine energies in states 1 and 2). The operator
\begin{equation}
\hat{V}=g\sum_{p,k}(\hat{b}_{k}^{+}\hat{a}_{k/2-p,2}\hat{a}_{k/2+p,1}+\hat
{a}_{k/2+p,1}^{+}\hat{a}_{k/2-p,2}^{+}\hat{b}_{k}) \label{13}%
\end{equation}
corresponds to the exchange resonance coupling between particles
in the open and closed channels. Hereafter, we suppose that volume
$\Omega=1$.

In the explicit form the coupling between pairs with zero total
momentum is selected from the nonresonance interaction between
atoms in the open channel in (\ref{1})
\begin{align}
\hat{V}_{a}^{\prime}  &  =U_{0}\sum_{p_{1},\,
p_{2}}\hat{a}_{p_{1},1}^{+}\hat
{a}_{-p_{1},2}^{+}a_{-p_{2},2}a_{p_{2},1},\label{012}\\
\hat{V}_{a}^{\prime\prime}  &  =U_{0}\sum_{p_{1},\, p_{2},\, q\neq0}\hat{a}%
_{p_{1},1}^{+}\hat{a}_{q-p_{1},2}^{+}a_{q-p_{2},2}a_{p_{2},1}. \label{12}%
\end{align}
The Hamiltonian in the form  (\ref{1}), in fact, coincides with
that widely used for the analysis of gases under Feshbach
resonance, see, eg.,
\cite{18},\cite{19},\cite{20},\cite{21},\cite{22}. The single
distinction is in the involvement of interaction (\ref{12}) which
role can be noticeable in the general case. Strictly speaking, in
Hamiltonian (\ref{1}) the terms with the nonresonance interaction
of atoms with quasi-molecules and  between quasi-molecules are
omitted. However, both these interactions are produced by the
resonance interaction  (\ref{13}).

The many-particle variational function is taken in the form
\begin{equation}
\Psi_{0}=\hat{\Psi}_{M}\hat{\Psi}_{M}^{\prime}\hat{\Psi}_{A}\left\vert
0\right\rangle
\label{6}%
\end{equation}
The field operator of atoms has a usual representation
\[
\hat{\Psi}_{A}=\prod\limits_{p}\left[
u_{p}+v_{p}\hat{a}_{p,1}^{+}\hat {a}_{-p,2}^{+}\right]
\]
In the coherent representation the field operator of
quasi-molecules in the state  $k=0$ has the form
\[
\hat{\Psi}_{M}=e^{-\frac{1}{2}M_{0}}\exp\left(  \sqrt{M_{0}}\hat{b}_{0}%
^{+}\right)
\]
It is assumed that the number of such quasi-molecules $M_{0}$ is
large.

A special feature of the structure for the many-particle
variational function (\ref{6}) is an involvement of the states
connected with virtual production of excited quasi-molecules. The
corresponding operator is represented as
\[
\hat{\Psi}_{M}^{\prime}=%
{\displaystyle\prod\limits_{k\neq0}}
[F_{k}+G_{k}\sum_{p}\hat{b}_{k}^{+}\hat{a}_{k/2-p,2}\hat{a}_{k/2+p,1}]
\]
Note that the order of operators in (\ref{6}) is essential.
Operator $\hat{\Psi }_{M}^{\prime}$ acts on the the state resulted
from the action of operator $\hat{\Psi}_{A}$\ on the vacuum. The
normalization of function (\ref{6}) results in two relations
\begin{equation}
u_{p}^{2}+v_{p}^{2}=1,\ F_{k}^{2}+w_{k}G_{k}^{2}=1, \label{10}%
\end{equation}
where
\begin{equation}
w_{k}=\sum_{p}v_{p+k/2}^{2}v_{p-k/2}^{2}. \label{011}%
\end{equation}
We assume that coefficients $u_{p},\ v_{p}$\ , $F_{k},\ G_{k}$ are
real.

The variational procedure reduces to finding the energy
\[
\bar{E}=\left\langle \Psi_{0}\left\vert
\hat{H}^{\prime}\right\vert \Psi _{0}\right\rangle
\]
and varying it over functions $v_{p}$\ , $F_{k},\ M_{0}$ with
account of relations (\ref{10}). Taking the structure of
variational function  (\ref{6}) and the form of Hamiltonian
(\ref{1}) into account,  we find
\begin{equation}%
\begin{array}
[c]{c}%
\bar{E}=2\sum_{p}\xi_{p}v_{p}^{2}+\zeta_{0}M_{0}+2\sqrt{M_{0}}\sum_{p}%
gu_{p}v_{p}\\
+\sum_{k\neq0}w_{k}(\zeta_{k}G_{k}^{2}+2gG_{k}F_{k})]\\
+U_{0}\left(  \sum_{p}u_{p}v_{p}\right)  ^{2}+U_{0}\left(  \sum_{p}v_{p}%
^{2}\right)  ^{2}.
\end{array}
\label{120}%
\end{equation}
For unit volume which we took, the magnitude $M_{0}$ equals, in
fact, to the density of quasi-molecules with $k=0$.

For the successive derivation of  (\ref{120}), functions
$v_{p}^{2}$ are multiplied by a factor $\exp(-z_{p})$ and
$u_{p}v_{p}$ by factor $\exp(-2z_{p})$ where $\
z_{p}=\sum_{k}G_{k}^{2}v_{p}^{2}v_{p-k}^{2}$. The direct
calculations show that magnitude $z_{p}$ proves to be small
compared with unity, at least, in the region of parameters
interesting for us. The involvement of these factors practically
has no influence on the results of the variational procedure. We
put them equal to 1.

Varying over $v_{p}$, we find
\begin{equation}%
\begin{array}
[c]{c}%
\frac{\delta\bar{E}}{\delta v_{p}}=4\xi_{p}v_{p}+\left(  u_{p}-\frac{v_{p}%
^{2}}{u_{p}}\right)  [2\sqrt{M_{0}}g+2U_{0}\sum_{p^{\prime}}u_{p^{\prime}%
}v_{p^{\prime}}]\\
+4U_{0}v_{p}\sum_{p^{\prime}}v_{p^{\prime}}^{2}\\
+\sum_{k\neq0}\frac{\partial w_{k}}{\partial v_{p}}(\zeta_{k}G_{k}^{2}%
+2gG_{k}F_{k}-gw_{k}\frac{G_{k}^{3}}{F_{k}})=0
\end{array}
\label{013}%
\end{equation}
Varying over $G_{k}$ and taking (\ref{10}) into account, one has
\begin{equation}
\frac{\delta\bar{E}}{\delta G_{k}}=\frac{2w_{k}}{F_{k}}\left[  \zeta_{k}%
G_{k}F_{k}+g-2gw_{k}G_{k}^{2}\right]  =0\label{14}%
\end{equation}
Using this equation and the relation resulted from  (\ref{011})
\[
\frac{\partial w_{k}}{\partial v_{p}}=4v_{p-k}^{2}v_{p},
\]
we rewrite (\ref{013}) in the form
\begin{equation}
2\xi_{p}^{\prime}u_{p}v_{p}+\left(  1-2v_{p}^{2}\right)  [\sqrt{M_{0}}%
g+U_{0}\sum_{p^{\prime}}u_{p^{\prime}}v_{p^{\prime}}]=0.\label{15}%
\end{equation}
Here
\begin{equation}
\xi_{p}^{\prime}=\xi_{p}+d_{p},\ d_{p}=\sum_{k\neq0}v_{p-k}^{2}gG_{k}%
F_{k}+U_{0}\sum_{p^{\prime}}v_{p^{\prime}}^{2}.\label{16}%
\end{equation}
Finally, varying over $M_{0}$, we arrive at
\begin{equation}
\zeta_{0}+g\frac{1}{\sqrt{M_{0}}}\sum_{p}u_{p}v_{p}=0.\label{17}%
\end{equation}
The extremum determined by conditions (\ref{14}), (\ref{15}), and
(\ref{17}) corresponds to the energy minimum provided that
\[
gu_{p}v_{p}<0,\ \ gG_{k}F_{k}<0
\]
Hence, in particular, it follows that the first term in (\ref{16})
and the second in  (\ref{17}) are negative. For definiteness,
henceforth we put  $g>0$.

Let us introduce the notation
\[
\Delta=g\sqrt{M_{0}}+U_{0}\sum_{p}u_{p}v_{p}%
\]
Solving equation (\ref{15}) with (\ref{10}), we find the
expressions similar to the BCS theory
\begin{equation}%
\begin{array}
[c]{c}%
v_{p}^{2}=\frac{1}{2}(1-\frac{\xi_{p}^{\prime}}{E_{p}}),u_{p}^{2}=\frac{1}%
{2}(1+\frac{\xi_{p}^{\prime}}{E_{p}})\\
2u_{p}v_{p}=-\frac{\Delta}{E_{p}},\ E_{p}=\sqrt{\xi_{p}^{\prime2}+\Delta^{2}}%
\end{array}
\label{20}%
\end{equation}
Then the equation for the gap $\Delta$ reads
\begin{equation}
\Delta+\frac{1}{2}U_{0}\sum_{p}\frac{\Delta}{E_{p}}=g\sqrt{M_{0}} \label{21}%
\end{equation}
At the same time equation (\ref{17}) goes over into
\begin{equation}
\varepsilon_{0}-2\mu=\frac{1}{2}g\frac{1}{\sqrt{M_{0}}}\sum_{p}\frac{\Delta
}{E_{p}} \label{22}%
\end{equation}
It is interesting that the role of virtual excited quasi-molecules
in (\ref{21}) and (\ref{22}) is reduced to the renormalization of
the atom spectrum, see, (\ref{16}) and (\ref{20}).

Solving equation (\ref{14}) with (\ref{10}) allows us directly to
find
\begin{equation}
G_{k}^{2}=\frac{1}{2w_{k}}(1-\sqrt{\frac{\zeta_{k}^{2}}{\zeta_{k}^{2}%
+4g^{2}w_{k}}}) \label{23}%
\end{equation}
Equations (\ref{21}) and (\ref{22}) contains formally an integral
diverging at the upper limit. This divergency is related with the
character of introducing parameters $\varepsilon_{0},\ g$ and
$U_{0}$ into Hamiltonian  (\ref{1}). In the general case this
requires the introduction an upper cutoff for the momentum region
region in which the interaction realizes effectively. However, in
some important cases such cutoff arises in the natural way.

Let us consider the most interesting case when the resonance
interaction $U_{0}=0$ is determining. Then from Eq. (\ref{21}) we
have
\begin{equation}
\Delta_{0}=g\sqrt{M_{0}} \label{24}%
\end{equation}
The magnitude $\varepsilon_{0}$ in Hamiltonian (\ref{1})
corresponds to the position of the resonance level. At $g\neq0$
the level renormalizes even in the limit of density tending to
zero, i.e., in vacuum. In this limit this can straightforwardly be
traced by calculating the scattering amplitude for a pair of atoms
with zero total momentum which is determined by interaction
(\ref{13}). Summing a series of the perturbation theory at
$E\rightarrow0$, we obtain, (cf. \cite{22}),
\[
f(E)=-\frac{\hbar
s/\sqrt{m}}{E-\varepsilon_{0}+\frac{1}{2}g^{2}\sum
_{p}\varepsilon{}_{p}^{-1}+is\sqrt{E}},\ s=\frac{m^{3/2}}{4\pi\hbar^{3}}g^{2}%
\]
It is seen that the resonance level energy at $g\neq0$
renormalizes and takes the value
\[
\tilde{\varepsilon}_{0}=\varepsilon_{0}-\frac{1}{2}g^{2}\sum_{p}%
\frac{1}{\varepsilon_{p}}%
\]
In the many-particle problem it is natural to treat magnitude
$\tilde{\varepsilon}_{0}$ as a real position of the resonance
level. Subtracting
$\frac{1}{2}g^{2}\sum_{p}\frac{1}{\varepsilon_{p}}$  from the both
sides of (\ref{22}) and taking (\ref{24}) into account, we have
\begin{equation}
\tilde{\varepsilon}_{0}-2\mu=\frac{1}{2}g^{2}\sum_{p}(\frac{1}{E_{p}}%
-\frac{1}{\varepsilon_{p}}) \label{26}%
\end{equation}
Provided that the nonresonance interaction is involved, the
situation becomes more complicated. However, treating (\ref{21})
and (\ref{22}) jointly, we find the equation which again does not
contain divergency
\begin{equation}
\tilde{\varepsilon}_{0}-2\mu=\frac{1}{2}g^{2}\sum_{p}(\frac{1}{E_{p}}%
-\frac{1}{\varepsilon_{p}})-\frac{g^{2}}{U_{0}}\frac{\left(
\Delta-\Delta
_{0}\right)  ^{2}}{\Delta\Delta_{0}} \label{26'}%
\end{equation}
If one employs the known relation  between  $U_{0}$ and
nonresonance scattering length $a_{bg}$ beyond the Born
approximation, see \cite{23}, the gap equation (\ref{21}) can be
transformed as
\begin{equation}
\Delta+\frac{1}{2}\Delta\frac{4\pi\hbar^{2}}{m}a_{bg}\sum_{p}(\frac{1}{E_{p}%
}-\frac{1}{\varepsilon_{p}})=g\sqrt{M_{0}} \label{21'}%
\end{equation}
For $g=0$, this equation reduces to the one familiar in the theory
of superconductivity, see \cite{13}. Assuming that the gas
parameter is small for the purely nonresonance interaction,
$U_{0}$ in (\ref{26'}) and (\ref{16}) can be replaced with
$4\pi\hbar^{2}a_{bg}/m$. Solving equations (\ref{26'}),
(\ref{21'}) and (\ref{23}) together with (\ref{10}), (\ref{16})
and (\ref{20}), we find values $\Delta,M_{0},G_{k}^{2}$ as a
function of chemical potential $\mu$ for fixed $g,$\ $a_{bg},$ and
$\tilde{\varepsilon}_{0}$. The values obtained allows us to
determine the number of atoms
\begin{equation}
N_{A}=\sum_{p,\sigma}\left\langle \Psi_{0}\right\vert \hat{a}_{p\sigma}%
^{+}\hat{a}_{p\sigma}\left\vert \Psi_{0}\right\rangle =2\sum_{p}v_{p}^{2}%
=\sum_{p}(1-\frac{\xi_{p}^{\prime}}{E_{p}}) \label{27}%
\end{equation}
and the number of virtually excited molecules
\begin{equation}
M^{\prime}=\sum_{k\neq0}\left\langle \Psi_{0}\right\vert \hat{b}_{k}^{+}%
\hat{b}_{k}\left\vert \Psi_{0}\right\rangle
=\sum_{k\neq0}w_{k}G_{k}^{2}
\label{28}%
\end{equation}
The relation for the total number of atoms in the system
\begin{equation}
N_{A}+2M_{0}+2M^{\prime}=N_{t} \label{29}%
\end{equation}
determines the equation to find chemical potential $\mu(N_{t})$.

As a result, the relative values $N_{A}/N_{t}$, $M_{0}/N_{t}$,
$M^{\prime}/N_{t}$ can be found as functions of
$\tilde{\varepsilon}_{0}/2\varepsilon_{t}$ for the fixed
dimensionless parameter
\begin{equation}
\gamma=\frac{g\sqrt{N_{t}}}{\varepsilon_{t}} \label{30}%
\end{equation}
Here $\varepsilon_{t}$ is the Fermi energy corresponding to the
total density $N_{t}$ of noninteracting atoms. It is worthwhile to
note that the system of equations obtained holds only for the
interval of variation $\varepsilon_{0}/2\varepsilon_{t}$ in which
condition  $M_{0}\gg1$ conserves.

Let us make an important remark. In the strongly interacting
many-particle system the superpositional character of the ground
state near the Feshbach resonance makes the definitions {\it atoms
and molecules} relatively conditional. In fact, this is
projections of the many-particle wave function onto the
corresponding states.

In the present paper we give a solution of the system of equations
obtained in some limiting cases which allow analytical
description. Let us first be $g\rightarrow0$ and $U_{0}>0$. From
Eq. (\ref{21'}) it follows that $\Delta=0$ and from  (\ref{23}) it
goes $G_{k}=0$.

According to (\ref{26'}) under these conditions the chemical
potential $\mu=\tilde {\varepsilon}_{0}/2$ for an arbitrary
$U_{0}\neq0$. The number of particles in the atom subsystem equals
\[
N_{A}=2%
{\displaystyle\sum\limits_{p}} \theta(-\xi_{p}^{\prime}).
\]
with quantity $\xi_{p}^{\prime}$ equal to
\[
\xi_{p}^{\prime}=\frac{p^{2}}{2m}-\frac{1}{2}\tilde{\varepsilon}_{0}%
+\frac{1}{2}U_{0}N_{A}%
\]
in accordance with (\ref{16}). Taking (\ref{29}) into account and
that $M_{g=0}^{\prime}=0$, we find straightforwardly that in the
interval
\begin{equation}
0<\frac{1}{2}\tilde{\varepsilon}_{0}<\varepsilon_{t}+\frac{1}{2}U_{0}N_{t}
\label{31}%
\end{equation}
gas of atoms in the normal state is in equilibrium with the
condensate of quasi-molecules. At the left-hand boundary at
$\tilde{\varepsilon}_{0}\rightarrow0$ the number of atoms
$N_{A}\rightarrow0$ and all particles prove to be in the
condensate of quasimolecules. As $\tilde{\varepsilon}_{0}$ grows,
the magnitude $M_{0}$ falls continuously. So, at the right-hand
boundary of interval (\ref{31}), $N_{A}=N_{t}$, i.e., the purely
atomic phase appears.  For $U_{0}<0$, the picture of behavior
$M_{0}$ and $N_{A}$ remains qualitatively same. However, the
condensate of quasi-molecules  vanishes with growth of
$tilde{\varepsilon}_{0}$ before, approximately at
$\frac{1}{2}\tilde{\varepsilon}_{0}\simeq\varepsilon_{t}
-\frac{1}{2}\left\vert U_{0}\right\vert N_{t}$. In fact, this
shift enhances to some extent owing to appearance of the gap
$\Delta$ in the spectrum of atoms, which can be found from solving
Eq. (\ref{21'}) at $g\rightarrow0$.

Let us turn to the solution of the system of equations obtained
above when the resonance interaction is dominant. We omit the
terms with $U_{0}$ and restrict ourselves with values $g$ which
correspond to the small dimensionless parameter $\gamma\ll1$. Then
in accordance with (\ref{24}) and (\ref{30}) one has
$\Delta/\varepsilon _{t} =\gamma\sqrt{M_{0}/N_{t}}\ll 1$.

One can straightforwardly see that  $\left\vert \ d_{p}\right\vert
\sim\gamma^{m}$ with $m\geq5/2$ in (\ref{16}) and the
renormalization of the spectrum can be neglected.  Let us start
from the consideration of interval
$\tilde{\varepsilon}_{0}\gg\Delta$. In this case from Eqs.
(\ref{26}) we can find
\begin{equation}
\mu=\frac{1}{2}\tilde{\varepsilon}_{0}-\frac{3}{8}\gamma^{2}\left(
\mu\varepsilon_{t}\right)  ^{1/2}\ln(\mu/\Delta) \label{32}%
\end{equation}
Calculating (\ref{27}), we have for the number of atoms
\[
\frac{N_{A}}{N_{t}}=\left(  \frac{\mu}{\varepsilon_{t}}\right)  ^{3/2}%
+O\left(  \frac{\Delta^{2}}{\mu^{2}}\right)
\]
Determining quantities $w_{k}$ (\ref{011}) and $G_{k}^{2}$
(\ref{23}), for the number of abovecondensate quasi-molecules
(\ref{28}) in the limit $\tilde{\varepsilon}_{0}\gg\Delta$ we
obtain
\[
\frac{2M^{\prime}}{N_{t}}\simeq\frac{1}{2}\gamma^{3/2}\left(
\frac{\mu
}{\varepsilon_{t}}\right)  ^{9/8}%
\]
Employing relation (\ref{29}), we find the number of condensate
quasi-molecules
\begin{equation}
\frac{2M_{0}}{N_{t}}\simeq1-\left(
\frac{\mu}{\varepsilon_{t}}\right)
^{3/2}-\frac{1}{2}\gamma^{3/2}\left(
\frac{\mu}{\varepsilon_{t}}\right)
^{9/8} \label{36}%
\end{equation}

The magnitude $\left(
\tilde{\varepsilon}_{0}/2\varepsilon_{t}\right) _{\ast}$, where
$M_{0}$ vanishes, one can estimate if the second term in
(\ref{32}) is omitted.
\begin{equation}
\left(  \frac{\tilde{\varepsilon}_{0}}{2\varepsilon_{t}}\right)  _{\ast}%
\simeq1-\frac{4}{3}\gamma^{3/2} \label{34}%
\end{equation}
In narrow interval of about $\gamma^{3/2}$ in the vicinity of this
boundary the number of excited quasi-molecules $M^{\prime}$ proves
to be larger than $M_{0}$. As  $\tilde{\varepsilon}_{0}$
decreases, the number of condensate particles $M_{0}$ increases
rapidly and $M_{0}\gg M^{\prime}$.

Let us find now the magnitude of sound velocity $C$ in the system.
For the equilibrium exchange of atoms between both subsystems, the
sound velocity is determined by the thermodynamic relation
\begin{equation}
C^{2}=\frac{N_{t}}{m}\frac{\partial\mu}{\partial N_{t}}, \label{35}%
\end{equation}
characteristic for the single-component systems, see, eg.,
\cite{23}. For $g\rightarrow 0$, one has $C^{2}=0$. This result is
a  consequence of  the transition of Fermi particles  with
compression to the quasi-molecular subsystem  and $\partial
p/\partial\rho=0$.

The similar effect takes place in an ideal Bose-gas at $T<T_{c}$
when a fraction of abovecondensate particles converts to the
condensate with the pressure growth.

When calculating the derivative $\partial\mu/\partial N_{t}$, only
the dependence $\Delta$ on $M_{0}$ in the logarithm argument
proves to be essential. (The involvement of the dependence of the
second term in (\ref{32}) on  $\mu$ gives corrections $\sim$
$\gamma^{2} $)
\[
\frac{\partial\mu}{\partial N_{t}}=\frac{3}{16}\gamma^{2}\left(
\mu\varepsilon_{t}\right)  ^{1/2}\frac{1}{M_{0}}\frac{\partial
M_{0}}{\partial
N_{t}}%
\]
Let us use the relation (\ref{36}) omitting the last term on the
right-hand side in the course of calculating $\partial
M_{0}/\partial N_{t}$. In the derivative there appears a term
proportional to $\frac{1}{M_{0}}\frac{\partial \mu}{\partial
N_{t}}$, which plays a significant role near the boundary
(\ref{34}) within the interval in which $M_{0}/N_{t}$ becomes
$\sim$ $\gamma^{2}$. Returning to definition (\ref{35}), we find a
general expression for the sound velocity
\begin{equation}
C^{2}=\frac{C_{\ast}^{2}}{1+\left(
\frac{\mu}{\varepsilon_{t}}\right)
^{1/2}\frac{3m}{2\varepsilon_{t}}C_{\ast}^{2}}\, ,\;\;\; C_{\ast}^{2}=\frac{3}{16}%
\gamma^{2}\frac{\varepsilon_{t}}{m}\left(
\frac{\mu}{\varepsilon_{t}}\right)
^{1/2}\frac{N_{t}}{M_{0}}%
\end{equation}
Far from the boundary (\ref{34})
\begin{equation}
C^{2}=C_{\ast}^{2} \label{801}%
\end{equation}
In the limit $\mu\rightarrow\varepsilon_{t}$ and
$M_{0}\rightarrow0$ we find
\begin{equation}
C^{2}=\frac{1}{3}\frac{2\varepsilon_{t}}{m} \label{0037}%
\end{equation}
It is interesting that this limiting value is reached from the
region where $M_{0}\neq0$  and the resonance interaction is
absent. Thus the sound velocity as a function of
$\tilde{\varepsilon}_{0}/2\varepsilon_{t}$ grows continuously to
value (\ref{0037}) typical for the collective oscillation of an
ensemble of free Fermi particles. For $\gamma^{2}\ll 1$, the
drastic growth $C^{2}$ to the value (\ref{0037}) occurs within
narrow interval of about $\gamma^{2}$.

Consider now the region $0<\tilde{\varepsilon}_{0}<\Delta$.
Equation (\ref{26}) in this limit goes over into
\begin{equation}
\mu\simeq\frac{1}{2}\tilde{\varepsilon}_{0}+0.8g^{2}\frac{\left(
2m\right)
^{3/2}}{4\pi^{2}}\Delta^{1/2} \label{370}%
\end{equation}
Accordingly, from (\ref{27}) we find
\[
\frac{N_{A}}{N_{t}}\simeq0.8\gamma^{3/2}\left(
\frac{M_{0}}{N_{t}}\right)
^{3/4}%
\]
Determining the value $M^{\prime}$ from  (\ref{28}), we have
\[
\frac{2M^{\prime}}{N_{t}}\simeq\frac{1}{2}\gamma^{21/8}%
\]
From the values obtained it follows directly that most of
particles in this region are in the condensate of quasi-molecules
\[
\frac{2M_{0}}{N_{t}}=1-\frac{1}{2}\gamma^{3/2}-\frac{1}{2}\gamma^{21/8}%
\]
Employing relations (\ref{35}) and (\ref{370}), we find for the
sound velocity
\begin{equation}
C^{2}\simeq\frac{\varepsilon_{t}}{3m}\gamma^{5/2} \label{38}%
\end{equation}
From  (\ref{370}) it is seen that the chemical potential tends to
zero at
\[
\tilde{\varepsilon}_{0}\simeq-\gamma^{5/2}\varepsilon_{t}%
\]
In order to obtain a complete picture, let us consider the region
of negative $\tilde {\varepsilon}_{0}$, assuming that
$\Delta\ll\left\vert \tilde{\varepsilon}_{0}\right\vert$. It is
evident that real molecules dominate in this region. Admixture of
atomic states is small.
\begin{equation}
\frac{N_{A}}{N_{t}}=\frac{3}{4}\gamma^{2}\left(  \frac{2\varepsilon_{t}%
}{\tilde{\varepsilon}_{0}}\right)  ^{1/2}\frac{M_{0}}{N_{t}} \label{39}%
\end{equation}

The number of abovecondensate molecules is small as well.
\[
\frac{2M^{\prime}}{N_{t}}\simeq\frac{1}{2}\gamma^{3}\left(
\frac{2\varepsilon
_{t}}{\tilde{\varepsilon}_{0}}\right)  ^{3/4}%
\]
Correspondingly, the number of molecules in the condensate is
\[
\frac{2M_{0}}{N_{t}}\simeq1-\frac{3}{8}\gamma^{2}\left(
\frac{2\varepsilon _{t}}{\tilde{\varepsilon}_{0}}\right)
^{1/2}-\frac{1}{2}\gamma^{3}\left(
\frac{2\varepsilon_{t}}{\tilde{\varepsilon}_{0}}\right)  ^{3/4}%
\]
Equation (\ref{26}) in this case takes the form
\[
\mu=\frac{1}{2}\varepsilon_{0}+0.55g^{2}\frac{m^{3/2}\left\vert
\mu\right\vert
^{1/2}}{\pi^{2}}+\frac{m^{3/2}}{32\pi\left\vert \tilde{\varepsilon}%
_{0}\right\vert ^{3/2}}g^{4}M_{0}.
\]
Hence, involving (\ref{30}), we find for the sound velocity
\begin{equation}
C^{2}\simeq\frac{3\pi}{256}\gamma^{4}\frac{\varepsilon_{t}}{m}\left(
\frac{2\varepsilon_{t}}{\left\vert
\tilde{\varepsilon}_{0}\right\vert
}\right)  ^{3/2}. \label{41}%
\end{equation}
The appearance of the high power of small parameter in (\ref{41})
is due to the interaction between molecules, which in this case
occurs via virtually created Fermi atoms. The number of the latter
atoms is small, see (\ref{39}). Accordingly,  $C^{2}$ tends to
zero at $\left\vert \tilde{\varepsilon}_{0}\right\vert
\rightarrow\infty$. Comparing (\ref{41}), (\ref{38}), (\ref{801}),
and (\ref{0037}), we see that the sound velocity increases
gradually with the growth $\tilde{\varepsilon}_{0}$, reaching its
maximum value (\ref{0037}).

Superfluidity and critical velocity $v_{c}$ in the system
considered are determined simultaneously by the spectra of
single-particle Fermi excitations and collective Bose excitations.
Accordingly, $v_{c}$ takes the minimum value from two velocities:
$C$ and $\min(E_{p}/p)$. Quantity $E_{p}$ is determined by
(\ref{20}). For $\tilde{\varepsilon}_{0}$ in the vicinity of
$2\varepsilon_{t}$, the critical velocity is governed by Fermi
excitations and $v_{c}\rightarrow0$ together with
$M_{0}\rightarrow 0$, $\Delta\rightarrow 0$, see (\ref{24}).
Velocity $\min(E_{p}/p)$ increases monotonously as
$\tilde{\varepsilon}_{0}$ decreases. On the contrary, the sound
velocity falls. The critical velocity therewith keeps to be
governed by single-particle excitations up to
$\tilde{\varepsilon}_{0}/2\varepsilon_{t}\simeq 1/3$. For smaller
values $\tilde{\varepsilon}_{0}/2\varepsilon_{t}$ including the
region of negative $\tilde{\varepsilon}_{0}$, the critical
velocity is determined by the sound velocity.

The work is supported by the Russian Foundation of Basic
Researches.


\bigskip

\begin{thebibliography}{9}                                                                                               %
\bibitem {1}E. Tiesinga, B.\,J. Verhaar, H.\,T.\,C. Stoof, Phys. Rev.
A\textbf{ 47}, 4114 (1993).

\bibitem {2}C.\,A. Regal, G. Ticknor, J.\,L. Bohn, D.\,S. Jin, Nature
\textbf{424}, 47 (2003).

\bibitem {3}K.\,E. Strecker, G.\,B. Partrige, R.\,G. Hulet, Phys. Rev.
Lett. \textbf{91}, 080406 (2003).

\bibitem {4}S. Jochim, \emph{et al.}, Phys. Rev. Lett. \textbf{91}, 240402 (2003).

\bibitem {5}J. Cubizolles, \emph{et al.}, Phys. Rev. Lett. \textbf{91},
240401 (2003).

\bibitem {6}D.\,S. Petrov, C. Solomon,
G.\,V. Shlyapnikov, Phys. Rev. Lett. \textbf{93}, 090404 (2004).

\bibitem {7}M. Greiner, C.\,A. Regal, D.\,S. Jin, Nature \textbf{426}, 537 (2003).

\bibitem {8}S. Jochim, \emph{et al.}, Science \textbf{302}, 2101 (2003).

\bibitem {9}M.\,W. Zwierlein, \emph{et al.}, Phys. Rev. Lett. \textbf{91},
250401 (2003).

\bibitem {10}T. Bourdel, \emph{et al.}, Phys. Rev. Lett. \textbf{93},
050401 (2004).

\bibitem {11}M. Holland, \emph{et al.}, Phys. Rev. Lett. \textbf{87},
120406 (2001).

\bibitem {12}E. Timmermans, \emph{et al.}, Phys. Lett. A \textbf{285}, 228 (2001).

\bibitem {13}A.\,J. Leggett, in Modern Trends in Theory of Condensed Matter,
p.13 (Springer-Verlag, Berlin 1980).

\bibitem {14}P. Nozi\`{e}res and S. Schmitt-Rink, J. Low. Temp. Phys.
\textbf{59}, 195 (1985).

\bibitem {15}C.\,A. Regal, M. Greiner, D.\,S. Jin, Phys. Rev. Lett.
\textbf{92}, 040403 (2004).

\bibitem {16}M.\,W. Zwierlein, \emph{et al.}, Phys. Rev. Lett. \textbf{92},
120403 (2004).

\bibitem {17}M.\,W. Zwierlein, \emph{et al.}, cond-mat/0412675 (2004).

\bibitem {18}Y. Ohashi, A. Griffin, Phys. Rev. A \textbf{67}, 063612 (2003);
Phys. Rev. Lett. \textbf{89}, 130402 (2002).

\bibitem {19}J. Stajic, \emph{et al.}, Phys. Rev. A \textbf{69}, 063610 (2004).

\bibitem {20}G.\,M. Bruun, C.\,J. Pethick, Phys. Rev. Lett. \textbf{92},
140404 (2004).

\bibitem {21}G.\,M. Falco, H.\,T.\,C. Stoof, Phys. Rev. Lett. \textbf{92},
130401 (2004).

\bibitem {22}A.\,V. Andreev, V. Gurarie, L. Radzihovsky, Phys. Rev. Lett.
\textbf{93}, 130402 (2004).

\bibitem {23}E.\,M. Lifshits, L.\,P. Pitaevskii, Statistical Physics, Part 2,
(Pergamon, London, 1980).
\end{thebibliography}
\end{document}